\documentclass[twocolumn,showpacs,prl,amsmath,amssymb,superscriptaddress]{revtex4}
\usepackage{color}
\usepackage{float}
\usepackage{verbatim}
\usepackage{calc}
\usepackage{float}
\usepackage{epsfig} 
\usepackage{subfigure} 
\usepackage{graphicx}% Include figure files
\usepackage{bm}% bold math
\def \equi#1{\mathrel{\mathop{\kern 0pt\sim}\limits_{#1}}} 
\scrollmode

\begin{document}
\title{Depletion-Controlled Starvation of a Diffusing Forager}
\author{O. B\'enichou}
\affiliation{Laboratorie de Physique Th\'eorique de la Mati\`ere Condens\'ee (UMR
  CNRS 7600), Universit\'e Pierre et Marie Curie, 4 Place Jussieu, 75255 Paris
Cedex France}

\author{S. Redner} \affiliation{Department of Physics, Boston University, Boston,
  Massachusetts 02215, USA and Santa Fe Institute, 1399 Hyde Park Road, Santa Fe, New
  Mexico 87501, USA}

\begin{abstract}

  We study the starvation of a lattice random walker in which each site
  initially contains one food unit and the walker can travel $\mathcal{S}$
  steps without food before starving.  When the walker encounters food, it is
  completely eaten, and the walker can again travel $\mathcal{S}$ steps
  without food before starving.  When the walker hits an empty site, the time
  until the walker starves decreases by 1.  In spatial dimension $d=1$, the
  average lifetime of the walker $\langle\tau\rangle\propto \mathcal{S}$,
  while for $d>2$, $\langle\tau\rangle\simeq\exp(\mathcal{S}^\omega)$, with
  $\omega\to 1$ as $d\to\infty$; the latter behavior suggests that the upper
  critical dimension is infinite.  In the marginal case of $d=2$,
  $\langle\tau\rangle\propto \mathcal{S}^z$, with $z\approx 2$.  Long-lived
  walks explore a highly ramified region so they always remain close to
  sources of food and the distribution of distinct sites visited does not
  obey single-parameter scaling.

\end{abstract}
\maketitle

Searching for a randomly located resource is an essential task of all living
organisms~\cite{C76,KR85,SK86,LKW88,OB90,B91,Vea96,ASD97,KM01,BLMV11}.  Examples
include searching for nourishment, an abode, or a particular individual.
Stochastically driven search processes also underlie diffusion-controlled
reactions~\cite{R85} and a variety of physiological processes~\cite{BB76}.
In all these examples, the time for a successful search is the typical metric
that the organism is trying to optimize.  A related aspect of stochastic
search is the tradeoff between continued exploitation of a familiar resource
or the exploration of new domains for potentially more fruitful
resources~\cite{GDB14}.

An important theme in biological foraging~\cite{C76,KR85,SK86} is the notion
of optimality.  Given an environment with heterogeneously distributed
resources and a predatory organism that has full environmental knowledge and
makes intelligent decisions, the marginal value theorem~\cite{C76} provides a
specific criterion about whether the forager should continue to exploit
current resources or move to new territory.  Our focus, in this Letter, is on
a different aspect of foraging in that we ascribe zero intelligence to the
forager and it is the \emph{depletion} of the environmental resource that
determines when the forager dies.  

In our model, the forager is a random walker that gradually depletes the
resource contained in a medium as it moves.  The medium is a $d$-dimensional
lattice with a unit of food initially at each site.  The walker is endowed
with an intrinsic starvation time $\mathcal{S}$, defined as the number of
steps it can take without encountering food before starving to death.  If the
walker encounters a food-containing site, the walker instantaneously and
completely consumes the food and can again travel $\mathcal{S}$ additional
steps without eating before starving.  Each time the walker encounters an
empty site, it comes one time unit closer to starvation.  Our goal is to
understand the interplay between the amount of the resource consumed and how
long the random walker can survive before starving.

We focus on two key observables of starving random walks: the average
lifetime $\langle\tau\rangle $ and the average number
$\langle\mathcal{N}\rangle$ of distinct sites visited when starvation occurs.
In dimension $d=1$, $\langle\tau\rangle\propto \mathcal{S}$ and we determine
the distribution of $\mathcal{N}$ at starvation, $P(\mathcal{N})$.  For
$d>2$, the transience of the random walk leads to $\tau$ scaling as
$\exp(\mathcal{S}^\omega)$, with $\omega\to 1$ as $d\to\infty$.  When
successive visits to new sites are uncorrelated, corresponding to $d=\infty$,
we find $\tau\sim e^{k\mathcal{S}}$, with $k$ given in terms of the
probability to visit a new site.  In $d=2$, numerical simulations suggest
that $\tau\simeq \mathcal{S}^z$, with $z\approx 2$.  We develop a mean-field
approximation for $d=2$ that gives a rigorous lower bound for $\tau$ and
suggests that $P(\mathcal{N})$ does not obey single-parameter scaling, as
seen in our simulations.

Our mortality mechanism differs from previous models in which a random walker
can die or be absorbed at a fixed rate, independent of its
location~\cite{T80,BNHW87,KF00,YAL13}.  Here, the lifetime distribution of a random
walker is not given {\it a priori} but is generated by the random-walk
trajectory, which renders the problem highly nontrivial.  It is worth
mentioning the related problem of the ``excited'' random walk, in which the
hopping of the walker depends on whether it has just encountered food or an
empty site~\cite{PW97,D99,BW03,ABV03,Z04,AR05}.  While the excited random
walk has surprising behavior, we will show that even when the motion of the
walker is not explicitly affected by the environment unusual properties
arise.

\begin{figure}[h]
\centerline{\includegraphics[width=0.4\textwidth]{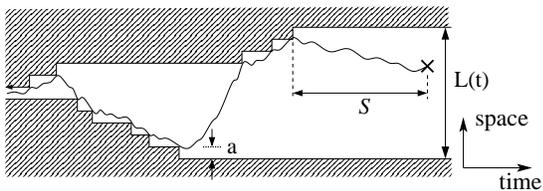}}
\caption{A $d=1$ starving random walker clears out an interval where food
  (shaded) has been eaten.  The walker starves ($\boldsymbol{\times}$) when
  it travels $\mathcal{S}$ steps without encountering food.}
\label{interval}  
\end{figure} 

{\it One dimension}.  As the walker moves, an interval devoid of food---a
desert---is gradually carved, and the survival of the walker is controlled by
the interplay between wandering within the desert and reaching food at the
edge of this desert (Fig.~\ref{interval}).  For a starving random walker to
survive for times beyond its intrinsic lifetime $\mathcal{S}$, excursions of
more than $\mathcal{S}$ steps without food cannot occur in its past history.
A long-lived walk must therefore spend less time wandering in the interior of
a desert than unrestricted walks.  Thus, the mean number of distinct sites
visited should be larger than that for unrestricted random walks of the same
number of steps $N$.

\begin{figure}[ht]
\centerline{\includegraphics[width=0.325\textwidth]{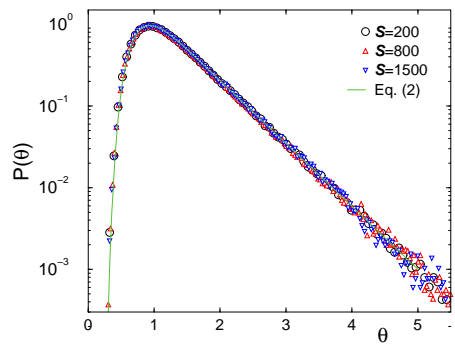}}
\caption{The scaled distribution of number of distinct sites visited for
  three representative values of $\mathcal{S}$.  The curve is the theoretical
  prediction from Eq.~\eqref{Q} and the data are based on $10^6$ walks for
  each value of $\mathcal{S}$. }
\label{s-1d-dist}  
\end{figure}

We first determine $P(\mathcal{N})$, the probability that a random walker has
visited $\mathcal{N}$ distinct sites when it starves.  This probability can
be expressed as
\begin{equation}
\label{PLdef}
P(\mathcal{N})= Q_2 \,Q_3\, Q_4 \ldots Q_\mathcal{N}(1-Q_{\mathcal{N}\!+1})\,,
\end{equation}
where $Q_j=\int_0^\mathcal{S} dt \, F_j(t)$, and $F_j(t)$ is the probability
that the walker reaches either end of an interval of length $ja$ (with $a$
the lattice spacing) at step $t$ when starting a distance $a$ from one end.
Each $Q_j$ accounts for the interval growing from length $j-1\to j $ because
the walker reaches either end point within $\mathcal{S}$ steps, while the
factor $Q_{\mathcal{N}+1}$ accounts for the last excursion in which the
walker starves.  It is expedient to express the visitation probability in
terms of the scaled number of distinct sites visited, $\theta =
a\mathcal{N}/(\pi\sqrt{D\mathcal{S}})$.  In the Supplemental Material, we
show that, when $\mathcal{S}\gg1$, $P(\theta)$ is (Fig.~\ref{s-1d-dist})
\begin{equation}
\label{Q}
P(\theta) \simeq \frac{4}{\theta}\!\sum_{m\geq 0} \!\!
\exp\Big\{\! -\frac{(2m\!+\!1)^2}{\theta^2} -2\!\sum_{n\geq 0} 
\mathrm{E}_1\!\Big[\frac{(2n\!+\!1)^2}{\theta^2}\Big]\!\Big\},
\end{equation}
where $D\equiv a^2/2$ is the diffusivity of the corresponding continuous
process and $E_1(x)\equiv\int_1^\infty \frac{dt}{t}\,e^{-xt}$ denotes the
exponential integral.  From Eq.~\eqref{Q}, the average number of visited
sites at the starvation time is
\begin{align}
  \langle \mathcal{N}\rangle \simeq
  \frac{\pi\sqrt{D\mathcal{S}}}{a}\int_0^\infty \theta \,P(\theta)\,
  d\theta\,\approx A\, \sqrt{\mathcal{S}}\,,
\end{align}
with $A\approx 2.902\,22$~\cite{S-compare}.

The average lifetime $\langle \tau\rangle$ of starving random walks is formally
given by
\begin{equation}
\langle\tau\rangle = \sum_{j\geq 1}
\Big(\langle\tau_1\rangle+\langle\tau_2\rangle+\dots +\langle\tau_j
\rangle+\mathcal{S}\Big) P(j)\,,
\end{equation}
where $\langle \tau_j\rangle$ is the average time for the random walk to hit
either end of the interval in the $j^{\rm th}$ excursion, conditioned on the
walker hitting either end before it starves, while the factor $\mathcal{S}$
accounts for the final excursion that causes the walker to starve.  By
definition
\begin{equation}
\label{tau-n}
  \langle \tau_j\rangle 
= \frac{\int_0^\mathcal{S} dt\, t \,F_j(t) }
{\int_0^\mathcal{S} dt\, F_j(t) } \,,
%  \langle \tau_j\rangle 
%= {\int_0^\mathcal{S} dt\, t \,F_j(t) }\,\,\bigg/
%{\int_0^\mathcal{S} dt\, F_j(t) } \,,
\end{equation}
The numerator, defined as $N_j$, reduces to
\begin{align}
N_j= \frac{4\pi D}{(ja)^2}\sum_{n\geq 0}
(2n\!+\!1)\sin \lambda_{j,n}\int_0^\mathcal{S} \!\!dt\,\, t\,\, e^{-\beta t}\,,
\end{align}
with $\lambda_{j,n}\equiv (2n\!+\!1)\pi/j$ and  $\beta\equiv \lambda_{j,n}^2/(ja)^2$.

\begin{figure}[ht]
\centerline{\includegraphics[width=0.325\textwidth]{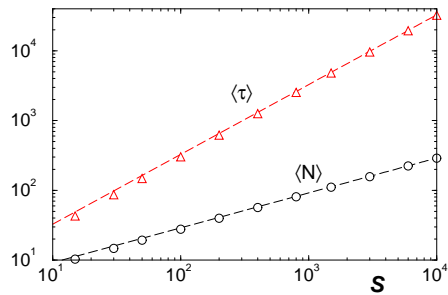}}
\caption{Average number of distinct sites visited $\langle N\rangle$
  ($\circ$) and average lifetime $\langle \tau\rangle$ $(\bigtriangleup$) for
  $10^6$ realizations of starving random walks in one dimension at the
  starvation time versus $\mathcal{S}$.  The dashed lines are the respective
  asymptotic predictions of $\langle \mathcal{N}\rangle\sim
  2.902\,22\sqrt{\mathcal{S}}$ and $\langle \tau\rangle\sim 3.267\,86 \ \mathcal{S}$. }
\label{st-1d}  
\end{figure}

For large $j$, we  approximate the sine function by its argument and
perform the temporal integral to give
\begin{align}
N_j
&  \simeq \frac{4a^2}{\pi^2 D}\sum_{n\geq 0} \frac{1}{(2n+1)^2}\left[1-
    e^{-\beta\mathcal{S}}(1+\beta\mathcal{S})\right]\,
\end{align}
for $\mathcal{S}\gg 1$.  Using this in Eq.~\eqref{tau-n} gives $\langle \tau
\rangle\simeq 3.26786\,\,\mathcal{S}$.  These results for the distinct sites
visited and the lifetime agree with numerical simulations shown in
Fig.~\ref{st-1d}.  \medskip

\begin{figure}[ht]
\centerline{\includegraphics[width=0.4\textwidth]{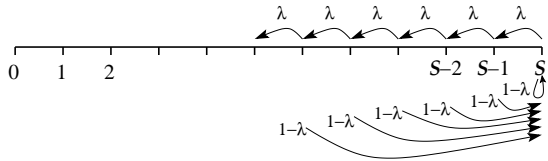}}
\caption{Evolution of a high-dimensional random walk in starvation space.  A
  particle at position $n$ in this space can survive $n$ additional steps
  without encountering food.}
\label{line}  
\end{figure}

{\it Infinite dimension.} For $d\!=\!\infty$, the probabilities of hitting a
previously visited or previously unvisited site equal $\lambda$ and
$1-\lambda$, respectively, with $\lambda$ a constant.  Schematically, the
time until the walker starves undergoes one-dimensional hopping in
``starvation space''---an interval of length $\mathcal{S}$---and starvation
occurs when $0$ is reached (Fig.~\ref{line}).  A particle at site $n$ in
starvation space can wander $n$ additional steps in physical space without
encountering food before starving.  When the walker hits a previously visited
site in physical space, the time to starvation decreases by one time unit,
corresponding to a hop to the left with probability $\lambda$ in starvation
space.  When the random walker encounters a new food-containing site, it can
wander an additional $\mathcal{S}$ steps until starvation occurs,
corresponding to a long-range rightward hop to site $\mathcal{S}$ in
starvation space with probability $1-\lambda$.

Using this equivalence to hopping in starvation space, we now compute $t_n$,
the average time until the walker starves when starting from site $n$.  These
starvation times satisfy the recursions~\cite{R01}
\begin{align}
t_n& = 1 +\lambda t_{n-1} + (1-\lambda) t_\mathcal{S}\hskip 0.35in
2\leq n \leq \mathcal{S}\,,\nonumber \\
t_{1}& = 1 +(1-\lambda)t_\mathcal{S}\,,
\end{align}
from which
\begin{equation}
\label{tS}
t_\mathcal{S}=\langle \tau\rangle= \frac{1}{\lambda^\mathcal{S}}\,\Big(\frac{1-\lambda^\mathcal{S}}{1-\lambda}\Big)~.
\end{equation}
The high-dimensional limit corresponds to $\lambda\to 0$, for which the
average starvation time grows exponentially with $\mathcal{S}$, in contrast
to the linear dependence for $d=1$.  

We may also obtain the distribution of the number of distinct sites visited
by the walker in physical space at the starvation time.  For this quantity,
we need the probability $R_n$ that the walk reaches $\mathcal{S}$ without
first hitting 0, when starting from site $n$ in starvation space.  Each such
return corresponds to the random walker visiting a new site in physical space
without starving.  These return probabilities satisfy the recursions
\begin{align}
R_n&= 1-\lambda + \lambda R_{n-1}    \hskip 0.35in  2\leq N\leq \mathcal{S}\,,\nonumber\\
R_1&=1-\lambda\,,
\end{align}
with solution $R_n=1-\lambda^n$. 

For a walker that starts at site $\mathcal{S}$ in starvation space, the
probability that $\mathcal{N}$ distinct sites are visited before the walker
starves is given by
\begin{equation}
P(\mathcal{N})= (R_\mathcal{S})^{\mathcal{N}}(1-R_\mathcal{S})\,,
\end{equation}
from which the average number of distinct sites visited before the walker
starves is
\begin{equation}
\langle \mathcal{N}\rangle = \frac{R_\mathcal{S}}{1-R_\mathcal{S}}= \lambda^{-\mathcal{S}}(1-
\lambda^{\mathcal{S}})\,.
\end{equation}
%which is the same as $\langle \tau\rangle$ from Eq.~\eqref{tS}, except for the
%factor of $1-\lambda$.

\begin{figure}[ht]
\centerline{\includegraphics[width=0.35\textwidth]{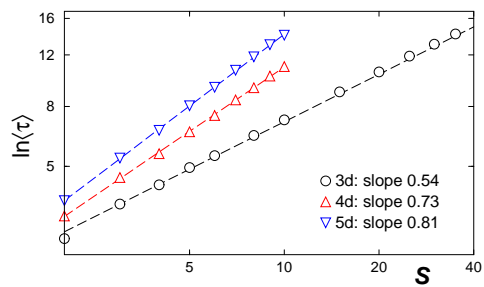}}
\caption{Logarithm of the average starvation time $\langle\tau\rangle$ versus
  $\mathcal{S}$ for starving random walks in $d=3$, 4, and 5.  Data are based
  on $10^6$ realizations for each value of $\mathcal{S}$.}
\label{tav-3d}  
\end{figure}

{\it Dimensions $d\!>\!2$.}  A random walk is transient, so new sites are visited
at a nonzero rate~\cite{W94,H95,R01} and it is unlikely for a walker to
first create a local desert and then wander within this desert until it
starves.  Thus the survival time should be much longer than in $d=1$ for the
same $\mathcal{S}$.  As a check, we simulate starving random walks on a
periodic hypercubic lattice where each site initially contains one food
unit.  We choose the lattice size so that wraparound effects are negligible.
%for the $\mathcal{S}$ values studied.  
A random walk dies when it takes $\mathcal{S}$ consecutive steps without
encountering food.  We find $\ln\langle\tau\rangle\sim \mathcal{S}^\omega$
with $\omega\approx 0.54$, $0.73$, and $0.81$ in $d\!=\!3,4,5$
(Fig.~\ref{tav-3d}).  Even though the walk is transient, so that the rate of
visiting new sites is nonzero, temporal correlations between successive
visits to new sites~\cite{future} lead to $\langle\tau\rangle$ deviating from
the mean-field result~\eqref{tS}.  The systematic trend in the exponent
estimates suggests that the upper critical dimension is $d_c\!=\!\infty$.

\begin{widetext}

\begin{figure}[ht]
\centerline{\includegraphics[width=0.275\textwidth]{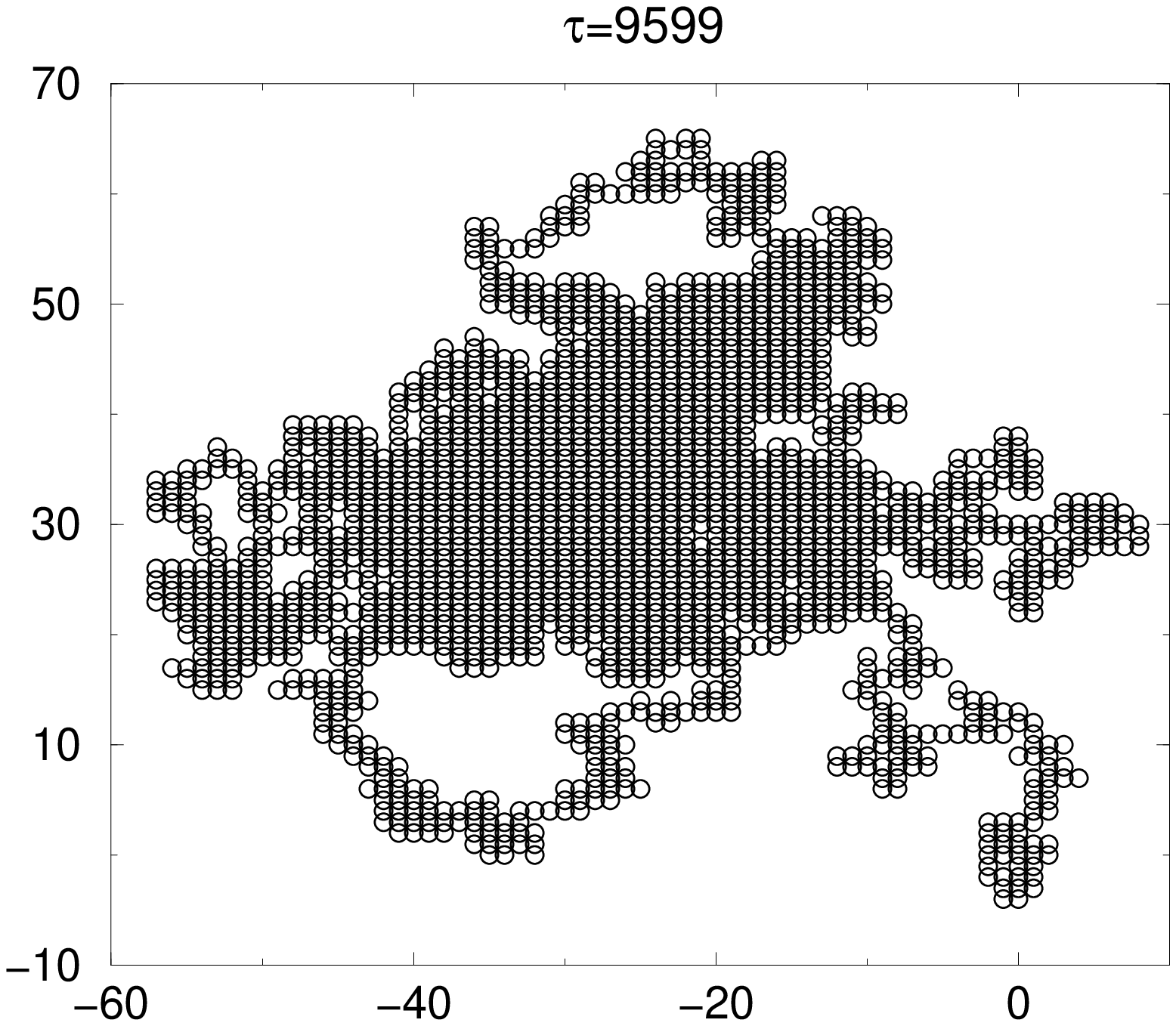}\qquad\qquad
\includegraphics[width=0.275\textwidth]{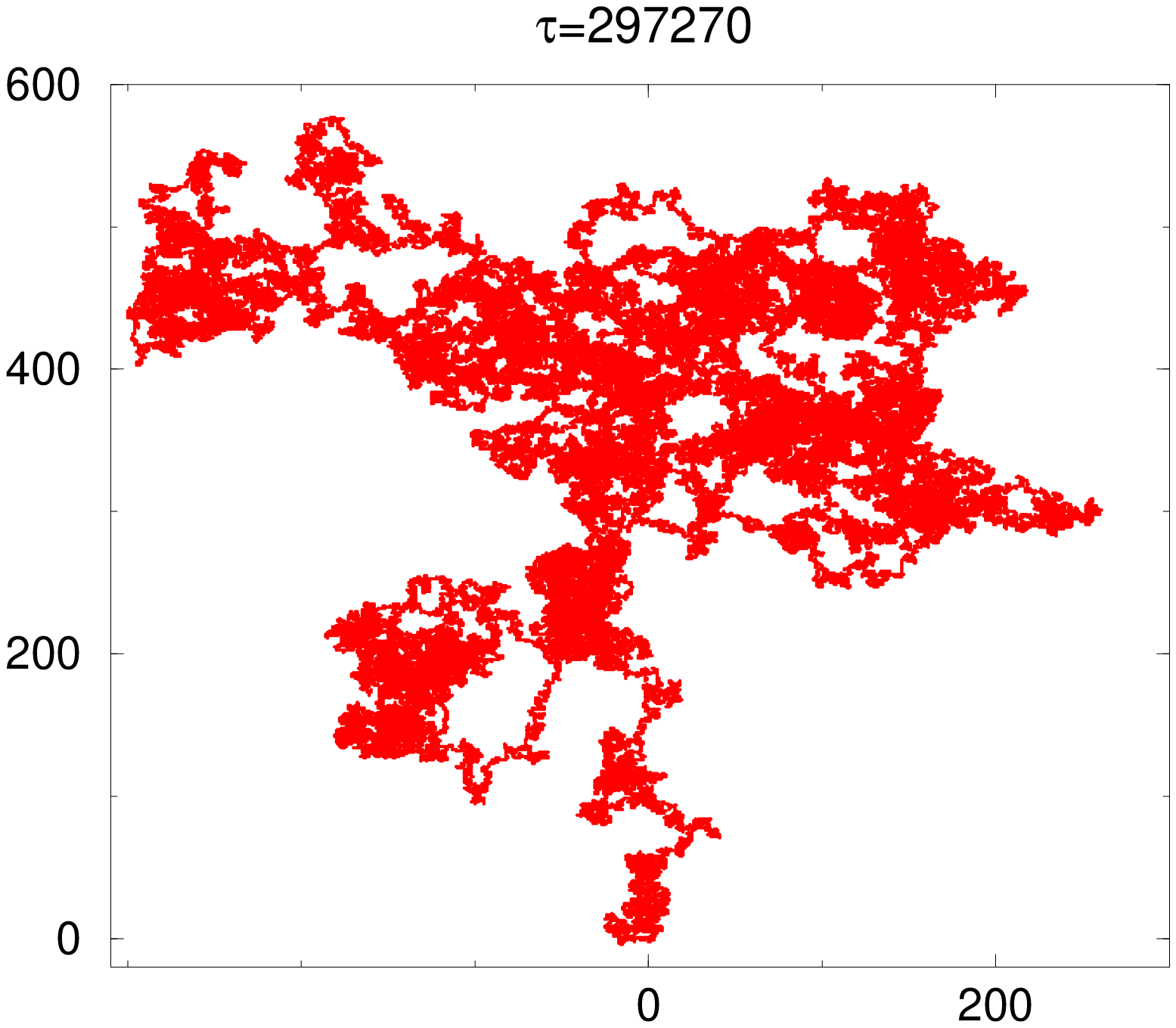}\qquad\qquad
\includegraphics[width=0.275\textwidth]{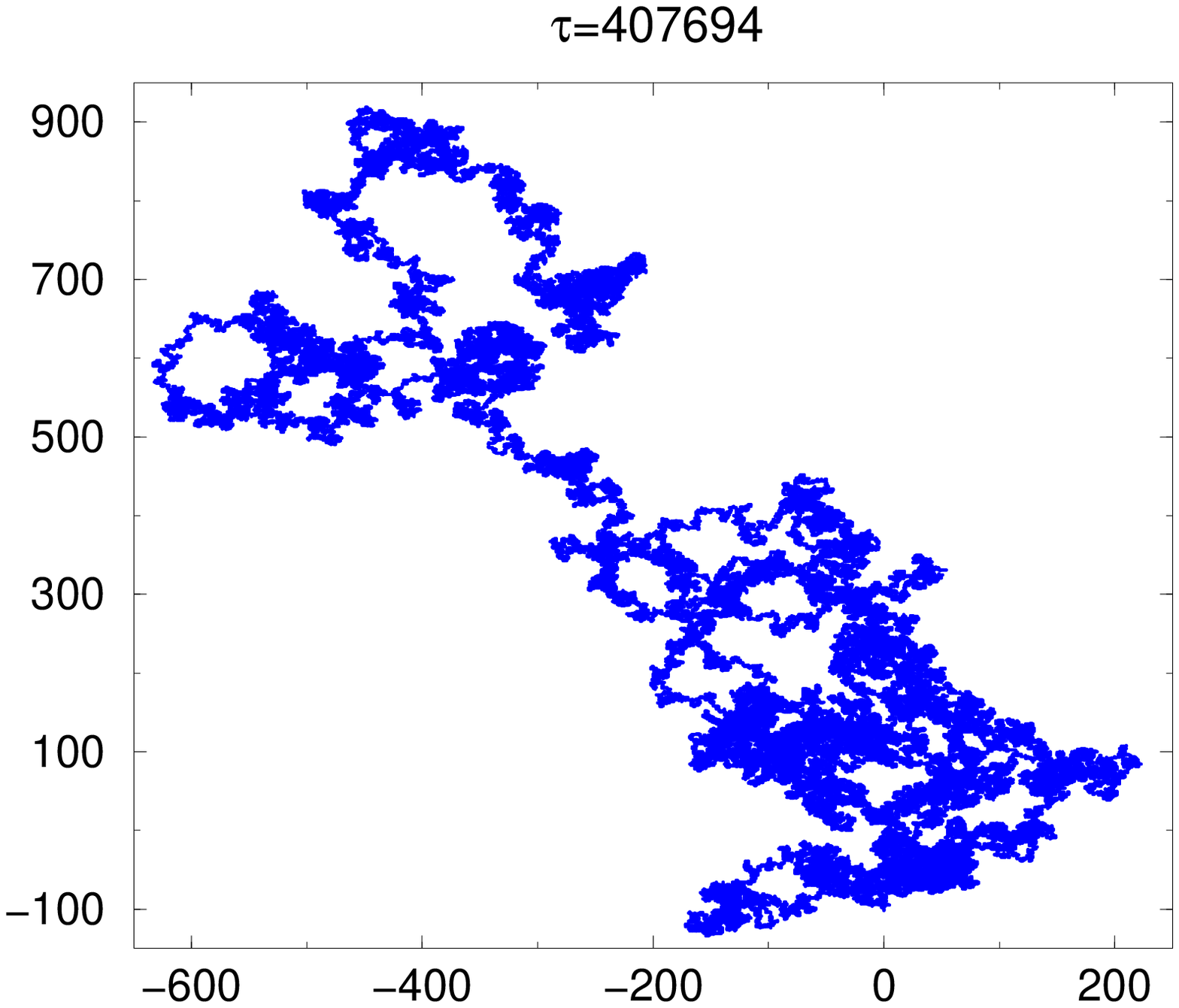}}
\caption{Example two-dimensional trajectories for $\mathcal{S}=500$.  The
  lifetime $\tau$ of each walk is indicated.  Note the very different spatial
  scales.}
\label{snapshots}
\end{figure}

\end{widetext}

{\it Two dimensions.} The most relevant case is that of two dimensions, which
lies at the boundary between recurrence and transience.  Numerical data
indicate that both $\langle\mathcal{N}\rangle$ and $\langle\tau\rangle$ scale
as $\mathcal{S}^2$, albeit with substantial preasymptotic corrections.  This
behavior is consistent with rigorous bounds for these
quantities~\cite{future}.  Sample random-walk trajectories are shown in
Fig.~\ref{snapshots} for intrinsic lifetime $\mathcal{S}=500$.  Trajectories
of short-lived walks are compact, while those of long-lived walks are quite
stringy so that the walker remains close to food-containing sites.

\begin{figure}[H]
\centerline{\includegraphics[width=0.35\textwidth]{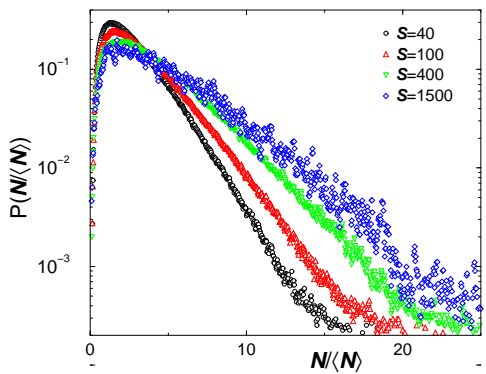}}
\caption{The scaled distribution of number of distinct sites visited for four
  representative values of $\mathcal{S}$.  The data have been averaged over a
  15-point range and only every fifth data point is displayed.  Data are
  based on $10^6$ realizations for each $\mathcal{S}$.}
\label{s-av-2d}  
\end{figure}

Strikingly, the underlying distribution for $\mathcal{N}$ violates
single-parameter scaling (Fig.~\ref{s-av-2d}).  To explain this feature, we
give a mean-field description for the evolution of the desert that relies on
the assumption that it always remains circular.  When the walker comes to the
edge of the desert and eats a morsel of food, we assume that the desert
becomes a slightly enlarged circle with an area that is increased by $a^2$.
After $n$ visits to the edge of the desert and immediately after the walker
has eaten, the radius of the desert is $R_n=a\sqrt{n/\pi}$ and the walker is
a distance $a$ from the edge.  Equation~\eqref{PLdef} still holds and $Q_j$
can be shown to be given by~\cite{future}
\begin{eqnarray}
Q_j=1-2\sum_{k=0}^\infty\frac{J_0\big(\sqrt{\pi/j}\alpha_k\big)}{\alpha_kJ_1(\alpha_k)}\,\,e^{-{D{\cal S}\pi\alpha_k^2}/{ja^2}},
\end{eqnarray}
where $\alpha_k$ is the $k^{th}$ zero of the Bessel function $J_0$.  We then
follow the same analysis as in one dimension to give~\cite{future} for $\mathcal{S}\gg 1$
\begin{align}
P({\cal N})&\simeq \exp\left\{\!-\frac{2D\pi{\cal S}}{a^2}\sum_{j=0}^\infty
\frac{\left[\theta e^{-\alpha_j^2/\theta}\!-\!\alpha_j^2{\rm E}_1(\alpha_j^2/\theta)\right]}{\alpha_jJ_1(\alpha_j)}\!\right\}\nonumber\\
&\hskip 0.5in \times 2\sum_{j=0}^\infty \frac{e^{-\alpha_j^2/\theta}}{\alpha_jJ_1(\alpha_j)}.
\end{align}
The simultaneous appearance of multiple parameters, $D$, $\mathcal{N}$, and
the scaling variable $\theta=a\mathcal{N}/(\pi\sqrt{D\mathcal{S}})$, shows
that the above expression for $P({\cal N})$ does not satisfy single-parameter
scaling, as is observed in our simulation shown in Fig.~\ref{s-av-2d}.

%This circular approximation also provides rigorous lower bounds for both the
%average number of distinct sites visited at starvation, $\langle \mathcal{N}
%\rangle \propto \mathcal{S}/\ln \mathcal{S}$, and the average lifetime,
%$\langle \tau \rangle \propto \mathcal{S}^{3/2}/ (\ln \mathcal{S})^3$.

To summarize, starving random walks represent a minimalist description for the
consumption of a depleting resource by a stochastic searcher.  The motion of
the walker is limited by the number of steps $\mathcal{S}$ that it can take
without encountering food before starving.  The spatial dimensionality plays
a crucial role in the dynamics, as the lifetime of a starving random walker
grows faster than algebraically in $\mathcal{S}$ for $d>2$ and algebraically
with $\mathcal{S}$ for $d\leq 2$.  We also obtained comprehensive results for
the starvation dynamics in $d=1$.  The two-dimensional case is particularly
challenging, as the distribution of the distinct sites visited does not obey
single-parameter scaling and the region visited by the random walker is
spatially complex.  Many challenges remain to understand all of the statistical
properties of this intriguing model, including the properties of starving
random walks at fixed time and the connection between the walk lifetimes and
the geometry of the desert, as illustrated in Fig.~\ref{snapshots}.

We acknowledge NSF Grant No.\ DMR-1205797 (S.R.) and ERC starting Grant
No. FPTOpt-277998 (O.B.) for partial support of this research.

\end{document}

\title{Supplementary Material for Depletion-Controlled Starvation of a Diffusing Forager}
\author{Olivier B\'enichou}
\affiliation{Laboratorie de Physique Th\'eorique de la Mati\`ere Condens\'ee (UMR
  CNRS 7600), Universit\'e Pierre et Marie Curie, 4 Place Jussieu, 75255 Paris
Cedex France}

\author{S. Redner} \affiliation{Department of Physics, Boston University, Boston,
  MA 02215, USA and Santa Fe Institute, 1399 Hyde Park Road, Santa Fe, New
  Mexico 87501, USA}
\maketitle

We wish to determine $P(\mathcal{N})$, the probability that a random walker
has visited $\mathcal{N}$ distinct sites when it starves.  This probability
is expressed as
\begin{equation}
\label{PLdef}
P(\mathcal{N})= Q_2 \,Q_3\, Q_4 \ldots Q_\mathcal{N}(1-Q_{\mathcal{N}\!+1})\,,
\end{equation}
where $Q_j=\int_0^\mathcal{S} dt \, F_j(t)$, and $F_j(t)$ is the probability
that the walker reaches either end of an interval of length $ja$ (with $a$
the lattice spacing) within $t$ steps when starting a distance $a$ from one
end.  Each $Q_j$ accounts for the interval growing from length $j-1\to j $
because the walker reaches either endpoint within $\mathcal{S}$ steps, while
the factor $Q_{\mathcal{N}+1}$ accounts for the last excursion in which the
walker starves.  It is expedient to express the the visitation probability in
terms of the scaled number of distinct sites visited, $\theta =
a\mathcal{N}/(\pi\sqrt{D\mathcal{S}})$.  We now provide the essential steps
to go from Eq.~\eqref{PLdef} to (2) in the manuscript.

In the long-time limit, we use the continuum expression for
$F_j(t)$~\cite{R01}:
\begin{eqnarray}
  F_j(t)=\frac{4\pi D}{(ja)^2}\sum_{n\geq 0} (2n+1)\, e_n(j/\sqrt{t})\, \sin\lambda_{j,n}\,,
\end{eqnarray}
where $D\equiv a^2/2$ is the diffusivity of the corresponding continuous
process, $\lambda_{j,n}\equiv (2n\!+\!1)\pi/j$, and 
\begin{equation}
e_n(j/\sqrt{t})\equiv
\exp\left[-(2n+1)^2\pi^2Dt/(ja)^2\right]\,.
\end{equation}
Consequently,
\begin{align}
Q_j  =\int_0^\mathcal{S} dt \, F_j(t)  = 1-\frac{4}{\pi}\sum_{n\geq 0} \frac{\,
  e_n(j/\sqrt{\mathcal{S}})\, \sin\lambda_{j,n}}{2n+1}\,.
\end{align}
This expression for $Q_j$, together with Eq.~\eqref{PLdef}, provides a formal
solution for the distribution of the number of distinct sites visited at the
starvation time.  

To obtain the explicit result, we start by taking the logarithm of the
function defined by
\begin{equation}
U_\mathcal{N}\equiv \prod_{2\leq j\leq \mathcal{N}} Q_j\,.
\end{equation}
This gives
\begin{equation}
  \ln U_\mathcal{N}= \sum_{2\leq j\leq \mathcal{N}} \ln\Big[ 1 -
  \frac{4}{\pi}\sum_{n\geq 0}\frac{ e_n(j/\sqrt{\mathcal{S}})\,\sin\lambda_{j,n}}{2n+1}\Big].
\end{equation}
We now convert the outer sum to an integral, introduce the variable $z\equiv
j/\sqrt{\mathcal{S}}$, and, for $\mathcal{S}\gg 1$, replace
$\sin\lambda_{j,n}/(2n+1)$ by $\pi/j$.  These steps give
\begin{equation}
\ln U_\mathcal{N} \simeq \sqrt{\mathcal{S}}\int_0^{\mathcal{N}/\sqrt{\mathcal{S}}}\!\!\! dz\, 
\ln\Big[1-\frac{4}{z\sqrt{\mathcal{S}}}\sum_{n\geq 0} 
e_n(z)\Big].
\end{equation}
Expanding the logarithm, which applies for large $\mathcal{S}$, gives
\begin{align}
\label{U}
\ln U_\mathcal{N} &\simeq - 4 \int_0^{\mathcal{N}/\sqrt{\mathcal{S}}} 
\frac{dz}{z}\sum_{n\geq 0} e_n(z),\nonumber \\
&= - 2\sum_{n\geq 0} \mathrm{E}_1\big[(2n+1)^2/\theta^2\big]\,,
\end{align}
where $\theta = a\mathcal{N}/(\pi\sqrt{D\mathcal{S}})$ is the scaled number
of distinct sites visited and $E_1$ is the exponential integral, which is
defined as $E_1(x)\equiv\int_1^\infty dt\,e^{-xt}/t$.

Similarly, the factor $1-Q_{\mathcal{N}\!+1}$ in \eqref{PLdef} has the
limiting behavior for $\mathcal{S}\gg 1$
\begin{equation}
\label{1-Q}
1-Q_{\mathcal{N}+1}\simeq \sum_{m\geq 0}
\frac{4a}{\pi\theta\sqrt{D\mathcal{S}}}\,\, e^{-(2n+1)^2/\theta^2}\,,
\end{equation}
Substituting Eqs.~\eqref{U} and \eqref{1-Q} in \eqref{PLdef} leads to Eq.~(2)
in the manuscript.